\documentclass[11pt]{article}
\usepackage{amsmath, amssymb, graphics,graphicx}
\usepackage{amscd}
\usepackage[dvips]{hyperref}
\usepackage{textcomp}

\title{Risk, VaR, CVaR and their associated Portfolio Optimizations when Asset Returns have a Multivariate Student T Distribution}

\author{William T. Shaw\thanks{Department of Mathematics and Computer Science, University College London. ({\tt w.shaw@ucl.ac.uk}).}}

\begin{document}

\maketitle

\begin{abstract}
We show how to reduce the problem of computing VaR and CVaR with Student T return distributions to evaluation of analytical functions of the moments. This allows an analysis of the risk properties of systems to be carefully attributed between choices of risk function (e.g. VaR vs CVaR); choice of return distribution (power law tail vs Gaussian) and choice of event frequency, for risk assessment. We exploit this to provide a simple method for portfolio optimization when the asset returns follow a standard multivariate T distribution. This may be used as a semi-analytical verification tool for more general optimizers, and for practical assessment of the impact of fat tails on asset allocation for shorter time horizons.\end{abstract}

\begin{center} 
\end{center}


\pagestyle{myheadings}
\thispagestyle{plain}
\markboth{W.T.SHAW}{W.T. Shaw: Risk, VaR and CVaR Portfolio Optimization with Student's T returns}

\section{Introduction}
The non-Gaussian nature of asset returns has been established for many years from many perspectives. From the point of view of statistical moment studies, Fama \cite{fama} and Mandelbrot \cite{mandel}  demonstrated the excess kurtosis of asset returns in the 1960s. Studies published in 2003 \cite{stanley} focused on tail analysis,  claiming that the tails of return distribution functions exhibit power-law behaviour, often with inverse {\it cubic} decay. Maximum Likelihood analysis has also been pursued. Fergusson and Platen \cite{platent} exhibited the importance of Student-T characteristics in {\it daily log-returns} of major indices. In the follow studies we will use the Student T as a {\it model} of asset returns, in that it can simultaneously reproduce excess kurtosis, power-law tails and be at or close to the MLE distributional estimate in a variety of situations\footnote{More recent work by Platen and co-workers \cite{epjb} finds the T or Variance Gamma as an MLE amongst the hyperbolic distributional family. My own preliminary studies also find Johnson-SU distributions as a contender, with all 3 types many orders of magnitude more likely than Gaussian on equity index and asset log-returns }. This is not to argue that this distribution is a universal panacea. Rather, it serves three other purposes. First, its manifestly greater likelihood in matching observed returns suggest that it is useful model for exploring risk characteristics and asset allocation in its own right. Second, it serves as a useful semi-analytical benchmark for more general computational models of optimization, to ensure their quality on an otherwise hard-to-obtain problem. VaR/CVaR optimization has been pioneered by Rockafeller and Uryasev \cite{cvaropt}, and general risk measure optimization by Shaw \cite{shawssrn}. Finally, the reduction of VaR and CVaR optimization to an essentially closed-form problem description, with an objective function that may be explicitly differentiated, opens the problem up to implementation on a wide variety of off-the-shelf optimizers. All of these studies and many others not cited here are attempting to bring greater diversity for choices of risk function to the pioneering work by Markowitz and collaborators, e.g. \cite{markow1, markow2, markow3}, and to build in explicit management of a variety of asset return distributions by functions beyond those that may be expressed simply in terms of the first two moments. 

The purpose of this note is not to engage in the already vigorous debate about VaR and CVaR as risk measures. There is a serious conceptual difficulty in estimating the probability of extreme events when there is so little data available. However, the view taken here is that several statistical studies over many decades based on the data that {\it is} available point to the Student T as a sensible model, and one that is many orders of magnitude more likely than a Gaussian model. In this sense the estimate of tail probabilities on such a basis is doing the {\it best one can given the data}, and is certainly more scientific than clinging to a Gaussian prescription and further indulging the pretence that unpredictable ``tsunami''-type events generate all the more extreme events.  A $25\sigma$ event is roughly $10^{130}$ times more likely in a $T_4$ model (as found in \cite{platent}) than it is in a Gaussian. There certainly are isolated severe shocks but the routine is also highly non-normal and capable of extreme movements.

\section{Recap of Gaussian VaR and CVaR}
Let $\phi(x) = (2\pi)^{-1/2}e^{-x^2/2}$ denote the PDF of the standard zero-mean and unit-variance normal distribution, and let $\Phi(x)$ denote the corresponding CDF with $\Phi' =\phi$. The Gaussian quantile function $Q_G(u)$ is the function satisfying the condition
\begin{equation}
\Phi(Q_G(u)) = u
\end{equation}
for all $u: 0<u<1$. This function is not available in simple closed-form, nor is its applied mathematics counterpart the inverse error function. However, its properties are well understood through the development of many numerical approximations \cite{wichura, ackweb} and more recently through its characterization as the solution of the non-linear ODE \cite{ejam}
\begin{equation}
Q'' = Q (Q')^2
\end{equation}
with boundary conditions $Q(1/2) = 0,\ Q'(1/2) = \sqrt{2\pi}$.

In the case of a Gaussian distribution with mean $\mu$ and variance $\sigma^2$, the quantile function of that distribution is then $\mu + \sigma Q_G(u)$ and the corresponding ``Value at Risk'', or VaR, is just the negative of the quantile, usually expressed in terms of a percentage number $w=100u$. That is
\begin{equation}
{\rm VaR}(w) = -\mu - \sigma Q_G(w/100)
\end{equation}
In what follows we will work in terms of the $u$-parametrization, with the function
\begin{equation}
{\rm VaR}(u) = -\mu - \sigma Q_G(u)
\end{equation}
So for example, the 2.5\% VaR is given by
\begin{equation}
 -\mu - \sigma Q_G(0.025)\sim -\mu +1.96 \sigma 
\end{equation}
and other choices of percentile lead to different weightings of the standard deviation.

The CVaR is a related entity, giving the {\it average} in a tail bounded by the corresponding VaR level . In general, for a density function $f(x)$ linked to a random variable $X$, we would need to work out
\begin{equation}
E[X|X \leq Q_G(u)] = \frac{1}{u} \int_{-\infty}^{Q_G(u)} x f(x) dx\ .
\end{equation}
In the case $f = \phi$, the integration may be done trivially, because $\phi' = -x \phi$, so that for a standard normal distribution
\begin{equation}
E[X|X \leq Q_G(u)] = -\frac{1}{u} \phi(Q_G(u))
\end{equation}
and applying a scaling and translation to get the general case gives us, for the {\it loss},
\begin{equation}
{\rm CVaR}(u) = -\mu + \frac{\sigma}{u} \phi(Q_G(u))
\end{equation}
In the case of a Gaussian system, there is no substantive difference between the VaR and CVaR measures: both take the form
\begin{equation}
{\rm Risk}(u) = -\mu + \psi(u) \sigma
\end{equation}
where $\psi_{GV}(u) = -Q_G(u)$ in the case of VaR and $\psi_{GC}(u) = \frac{1}{u}\phi(Q_G(u))$ in the case of $CVaR$.  And the numerical values are very similar.
\begin{table}[hbt]
\begin{center}
\begin{tabular}{|l|l|l|l|} \hline
quantile &  $\psi_{GV}$ & $\psi_{GC}$ &Time scale \\ 
$u$ &  $-Q_G(u)$ & $\phi(Q_G(u))/u$& (approx) \\ 
&&& \\ \hline
$0.100$  &$1.28155$ & $1.75498$& Fortnight \\ \hline
$0.025$  &$1.95996$ & $2.33780$& Two months\\ \hline
$0.010$  &$2.32635$  & $2.66521$&$~ 5$ months\\ \hline
$10^{-4}$  &$3.71902$& $3.95848$& 40Y\\ \hline
$10^{-6}$  &$4.75342$ &$4.94833$& 4000Y\\ \hline
\end{tabular}
\end{center}
\caption{$\sigma$-multipliers for Gaussian VaR and CVaR}
\end{table}
The numbers in Table 1 illustrate the impotence of switching from VaR to CVaR while remaining in a Gaussian framework. There is essentially no change in the practical outcome of the risk computation, despite the improved mathematical properties of CVaR with regard to coherence. One presumes this effect is well-known. 

\subsection{Recap of Gaussian VaR and CVaR Portfolio Optimization}
If the risk is now associated with the returns on a portfolio where the returns on $N$ assets follow a multivariate Gaussian distribution, then if the asset weights are $w_i$, $i=1,\dots,N$, the expected returns are $\mu_i$, and the covariance is $C_{ij}$, then the risk function may now be written as

\begin{equation}
{\rm Risk}(u, w_i) = -\sum_{i=1}^N\mu_i w_i + \psi(u) \sqrt{\sum_{i,j}C_{ij}w_i w_j}
\end{equation}
So both the VaR and CVaR optimizations are of the form of a mean-standard deviation optimization where the function $\psi$ characterizes the risk-aversity. This equation may be written down without further discussion because if $X_i$ has a multivariate normal distribution then $\sum_{i=1}^N w_i X_i$ has a univariate normal distribution for any collection of weights $w_i$. Supplemental constraints may be written down in the normal way - for a long-only fully-invested portfolio we would have the simple set
\begin{equation}
w_i \geq 0 ,\ \ \sum_{i=1}^N w_1 = 1\ .
\end{equation}
and in general we might relax these to allow short-selling and/or add further matrix conditions or weight inequalities.

Even in the Gaussian case we also see that optimizing VaR and optimizing CVaR are {\it different computations} and wil have differing optimal weights, except in the case where the mean may be ignored, for example in the limit $u \rightarrow 0$, when the problem is pure variance minimization in either case. 

It should be clear that in a real-world context where such a Gaussian model might, where it is appropriate at all, be more properly applied to the logs of asset returns, we are assuming that the time period of interest is short enough such that the approximation $e^R \sim 1+R$ may be used. This will also apply to the T model to be developed next. While this limits the general practical usefulness of Gaussian or Student T assumptions within a detailed optimization, it does not diminish the usefulness of a semi-analytical model for checking more general numerical techniques, and nor does it diminish the importance of getting a quantitative appreciation of the impact of fat tails on both the risk measures and the optimizations.

\section{VaR/CVaR for non-Gaussian models}
It will be clear from the discussion above that having a simple analytical model of the risk and portfolio optimization problems, at least of the specific type given previously, relies on the following properties in order to have a degree of tractability:
\begin{enumerate}
\item Having a finite variance;
\item Linear closure of the distributional family for the returns;
\item Having a formula for the quantile (for the VaR);
\item Being able to integrate $x*f(x)$ (for the CVaR).
\end{enumerate}
We will now elaborate on the satisfaction of these tractability conditions for the Student T. This is really just a matter of assembling known facts and characterizing the distribution in the right way. 

\subsection{Tractability of the Student T risk}
Student's T distribution was introduced by William Sealy Gosset in 1908 \cite{student} in connection with the statistics of small samples. In Gosset's original construction, expressed in a modern notation, the T random variable is constructed from $n+1$ (integer) standard independent normal random variables $Z_0, Z_1,\dots, Z_n$ as
\begin{equation}
T = Z_0 (\frac{1}{n}\sum_{k=1}^n Z_i^2)^{-1/2}
\end{equation}
Rewriting this in terms of a chi-squared denominator of degree $n$ - the `degrees of freedom', we have
\begin{equation}
T = Z_0 (\frac{1}{n}\chi_n^2)^{-1/2}
\end{equation}
and the condition that $n$ be an integer can then be relaxed, and we can consider real degrees of freedom in terms of a positive real variable $\nu$.
\begin{equation}
T = Z_0 (\frac{1}{\nu}\chi_{\nu}^2)^{-1/2}
\end{equation}
The density of $T$ may be derived by conditioning on the possible values of the denominator, and an elementary computation (see e.g. \cite{shawjcf}) leads to the probability density function
\begin{equation}
h(t,\nu) = \frac{1}{\sqrt{\nu \pi}}\frac{\Gamma[(\nu+1)/2]}{\Gamma[\nu/2]}\frac{1}{(1+t^2/\nu)^{(\nu+1)/2}}
\end{equation}
When $\nu >0$ the variance of this distribution exists provide $\nu>2$ and is then
\begin{equation}
V(T) = \frac{\nu}{\nu-2}
\end{equation}
In the multivariate case there are many possible extensions of the T to consider - see e.g. Shaw and Lee \cite{jmva} for an extensive discussion of the bivariate case, allowing for independent or non-independent marginals, and differing degrees of freedom in the marginals. However, for tractability of the risk problem we appear to be limited to the ``standard'' multivariate T where the $T_i$ are given by\footnote{If the reader wonders why the usual multivariate density is not being given, the reason is that the covariance and linear closure properties are much more transparent in this conditionally Gaussian form, which arises in simulation of the distribution and the associated copula.}
\begin{equation}
T_i  = (\chi_{\nu}^2/\nu)^{-1/2}
\sum_{k=1}^N A_{ik}Z_k \ .\end{equation}
for $N$ independent standard Gaussian random variables $Z_j$ and a mixing matrix $A_{ij}$. The covariance of this system is then just
\begin{equation}
E[T_i T_j] = \frac{\nu}{\nu-2} (A.A^T)_{ij}
\end{equation}
and if we have weights $w_i$, the random variable
\begin{equation}
\tilde{T} = \sum_{i=1}^N w_i T_i
\end{equation}
also has a T-distribution with variance
\begin{equation}
V(\tilde{T}) = \frac{\nu}{\nu-2} \sum_{ij}w_i (A.A^T)_{ij} w_j
\end{equation}
So in this case we manifestly have the {\it requisite linear closure} and simple expressions for the variance. 

\subsection{Do we need finite variance?}
Taleb \cite{taleb} has argued that ``finiteness of variance is irrelevant in the practice of quantitative finance'', in a paper so titled. I should point out that I mostly agree with this. The Student T distribution is perfectly well defined in the domain $1 < \nu \leq 2$ where the mean exists but the variance does not. In the domain $1 < \nu \leq 4$ the kurtosis is already theoretically infinite and there is no reason of principle why we cannot consider infinite variance within this family either. What {\it is} pathological about the case $\nu \leq 2$ is that the paramatrization adopted here and the reduction to a mean-sd optimization will break down. We need a suitable $L^1$ model of the dispersion to employ and in terms of which to express the VaR and CVaR. This will be explored elsewhere, and for this paper we will remain in the conservative domain where the underlying variance exists. 

Within this conservative domain the real issue is whether we have a meaningful and sensible positive definite covariance, and all the usual caveats apply. Our historical estimates may be poor for the future for both volatilities and correlations, and over-writing such data with estimates from e.g. options or traders is fraught with conceptual and technical problems. The use of a what-if approach with correlations, or dealing with many dimensions, may easily produce non positive-definite structures, and this needs to be properly analysed and managed. 

\subsection{VaR and CVaR in the T model}
In this case we are able to follow a similar pattern to that defined for the Gaussian case, except that we need to be careful with the variance normalization and manage the quantile structure. Suppose for now that we can find an expression for the quantile function $Q_T(u, \nu)$ for a standard univariate T distribution. If we wish to parametrize the problem again by mean and standard deviation, the VaR will be given by
\begin{equation}
{\rm VaR}(u) = -\mu - \sigma \sqrt{\frac{\nu-2}{\nu}}Q_T(u, \nu)
\end{equation}
so that the relevant $\psi$ function for this case is not the (negative of the) standard T quantile but the negative of the scaled {\it unit variance} T quantile\footnote{This would of course require modification to extend to the domain $1 <\nu \leq 2$.} $\tilde{Q}$:
\begin{equation}
\psi_{TV}(u) = -\tilde{Q}_T(u,\nu) = -\sqrt{\frac{\nu-2}{\nu}}Q_T(u, \nu)
\end{equation}
The VaR measure of risk is once again in standard form $-\mu+\psi_{TV}(u)\sigma$.

For the CVaR we no longer have an identity as simple as $\phi' = -x \phi$ that allowed the immediate integration of the Gaussian expressions, but it is a simple matter to find the integral of $-t h(t,\nu)$, since the density just contains powers of a linear function of $t^2$. On doing some elementary integration we find that the function
\begin{equation}
k(t,\nu) = \frac{\nu ^{\nu /2} \Gamma \left(\frac{\nu -1}{2}\right) \left(\nu
   +t^2\right)^{\frac{1}{2}-\frac{\nu }{2}}}{2 \sqrt{\pi } \Gamma \left(\frac{\nu
   }{2}\right)}
\end{equation}
is the correct expression, and the CVaR for a distribution with mean $\mu$ and variance $\sigma^2$ is now
\begin{equation}
CVaR(u) = -\mu + \sigma \sqrt{\frac{\nu-2}{\nu}}k(Q_T(u, \nu),u)\frac{1}{u}
\end{equation}
so that the $\psi$-function is now
\begin{equation}
\psi_{TC}(u) = \sqrt{\frac{\nu-2}{\nu}}k(Q_T(u, \nu),\nu)\frac{1}{u}
\end{equation}
Apart from the characterization of the Student T quantile function, we now have explicit expressions for VaR and CVaR for this distribution that are comparable with those that are already clear for the Gaussian case. 
\section{Characterizing the Student T quantile}
The inverse CDF, of quantile function, for the Student T was first\footnote{So far as this author is aware.} studied by Hill in his short and beautiful 1970 paper \cite{hill}. I revisited the matter in \cite{shawjcf} and exhibited the simple closed forms available for $\nu=1,2,4$, with the two simple even cases arising as the available characterization as the solution for even $\nu$ in terms of a polynomial of degree $\nu-1$. The case $\nu=1$ has been known for a long time as it is the Cauchy distribution quantile:
\begin{equation}
Q(u,1) = \tan \left[ \pi (u-1/2) \right]
\end{equation}
The case $\nu=2$ was known to Hill and takes the form, in our notation,
\begin{equation}
Q(u,2) = \frac{2u-1}{\sqrt{2u(1-u)}}
\end{equation}
Of course, neither of these older forms have finite variance. The only simple closed-form case with this property was derived in \cite{shawjcf} for $\nu = 4$ and is given by
\begin{equation}
\begin{split}
\alpha &= 4u(1-u)\\
q &=\frac{4}{\sqrt{\alpha}} \cos \left(\frac{1}{3} \cos^{-1}(\sqrt{\alpha})  \right)\\
Q(u,4) &= {\rm sgn}(u-1/2) \sqrt{q-4}
\end{split}
\end{equation}
As usual $\text{sgn}(x)$ is $+1$ if $x>0$ and $-1$ if $x<0$. In this case the relevant unit variance function is
\begin{equation}
\psi_{TV}(u) = -\frac{1}{\sqrt{2}} Q(u,4)
\end{equation}
\subsection{Inverse Beta Function representation}
The CDF for general positive real $\nu$ may be written in terms of $\beta$-functions:
\begin{equation}F_{\nu}(x) = \frac{1}{2}\left(1 + \text{sgn}(x) (1 - I_{(\frac{\nu}{x^2+\nu})}\left(\frac{\nu}{2},\frac{1}{2}\right)\right)\end{equation}
giving an expression in terms of regularized $\beta$-functions.  The regularized beta function $I_x(a,b)$ employed here is given by
\begin{equation}I_x(a,b) = \frac{B_x(a,b)}{B(a,b)}\end{equation}
where $B(a,b)$ is the ordinary $\beta$-function and $B_x(a,b)$ is the incomplete form
\begin{equation}B_x(a,b) = \int_0^x t^{(a-1)} (1-t)^{(b-1)} dt\end{equation}
Having got such a representation, this may be formally inverted to give the formula for the quantile:
\begin{equation}Q(u,{\nu}) = \text{sgn}\left(u-\frac{1}{2}\right)
\sqrt{\nu \left(\frac{1}{I_{\text{If}\left[u<\frac{1}{2},2 u,2 (1-u)\right]}^{-1}\left(\frac{\nu}{2},\frac{1}{2}\right)}-1\right)} \end{equation}
This last result is useful in computer environments where representations of inverse beta functions are available. 
If only the (forward) $\beta$-function is available, 
one can make the inversion by employing a stepping or bisection approach. For checking we give a {\it Mathematica} implementation in the form:
\begin{verbatim}
Q[u_, n_] := Module[{arg = If[u < 1/2, 2 u, 2 (1 - u)]},
  Sign[u - 1/2] Sqrt[n*(1/InverseBetaRegularized[arg, n/2, 1/2] - 1)]]
\end{verbatim}
This is not fast enough for live Monte Carlo simulation by the inverse transform method but gives high-precision quantile numbers. These need to be multiplied by $-\sqrt{(\nu-2)/\nu}$ to obtain the unit-variance loss $\psi$-function.
\subsection{Tail Series}
While for Monte Carlo simulation by inversion one needs rapidly-computable expressions for the entire interval $0<u<1$, with identified single- or double-precision accuracy, for our purposes here we just need a good approximation for the region $0<u\leq 0.025$ in order to evaluate VaR/CVaR values at the 2.5\% or less frequency levels. The following tail series is good enough to produce precision of better than $10^{-3}$ for $2 \leq \nu \leq 11$ and $0<u<0.025$:
\begin{equation}
\begin{split}
w &= \left(u\  \nu \sqrt{\pi}\Gamma[\nu/2]/\Gamma[(\nu+1)/2]  \right)^{2/\nu}\\
\beta &= \sum_{k=1}^{6} d_k w^k\\
Q(u,\nu) &= -\sqrt{\nu*(1/\beta-1)}
\end{split}
\end{equation}
The coefficients $d_k$ are given in Appendix A. The series is exact for $\nu=2$ and the precision for $2\leq\nu\leq4$ is better than $10^{-5}$.  The derivation of this and the associated recurrence for making more terms is given in Section 3.4 of \cite{ejam}. The final normalization to unit variance is also required to obtain $\psi_{TV}$.
\subsection{Sample numbers}
Practitioners will perhaps find sample values of the loss functions $\psi$ more illuminating, and such precomputed values may of course be fed into an off-the-shelf optimizer. In Table 2 we report the zero mean loss functions $\psi$ as a multiple of the standard deviation. The table has some striking numbers in it. First note the general similarity of the levels of the 2.5\% values for either VaR or CVaR for all the distributions - most are in the $2-3\sigma$ range. One can even see that the numbers {\it decrease} at very low powers. This is an odd effect but is correct due to the shape of the density. A casual analysis of losses within such 40-day windows might well lead one to conclude that a Gaussian model is enough or even conservative. The $T_4$ 2.5\% quantile is almost exactly the same as the Gaussian value! 

\begin{table}[hbt]
\begin{center}
\begin{tabular}{|l|l|l|l|l|l|} \hline
 &  & \multicolumn{4}{|c|}{Quantile Measures}  \\ \hline
 &Time $\sim$ & 2M & 5M & 4Y & 40Y \\
 DOF& $u = $& $0.025$ & $0.01$ & $10^{-3}$ & $10^{-4}$\\ \hline \hline
$\nu=\infty$  & $\psi_{GVaR}$& $1.96$ & $2.33$ & $3.09$ & $3.72$\\
(Gauss)                    & $\psi_{GCVaR}$& $2.34$ & $2.67$ & $3.37$ & $3.96$\\ \hline \hline
                    $\nu=6$  & $\psi_{TVaR}$& $2.00$ & $2.57$ & $4.25$ & $6.55$\\
              & $\psi_{TCVaR}$& $2.66$ & $3.29$ & $5.24$ & $7.95$\\ \hline \hline
$\nu=5$  & $\psi_{TVaR}$& $1.99$ & $2.61$ & $4.57$ & $7.50$\\
  & $\psi_{TCVaR}$& $2.73$ & $3.45$ & $5.82$ & $9.44$\\ \hline \hline
$\nu=4$  & $\psi_{TVaR}$& $1.96$ & $2.65$ & $5.07$ & $9.22$\\
  & $\psi_{TCVaR}$& $2.82$ & $3.69$ & $6.85$ & $12.3$\\ \hline \hline
$\nu=3$  & $\psi_{TVaR}$& $1.84$ & $2.62$ & $5.90$ & $12.8$\\
  & $\psi_{TCVaR}$& $2.91$ & $4.04$ & $8.90$ & $19.3$\\ \hline \hline
$\nu=2.5$  & $\psi_{TVaR}$& $1.60$ & $2.39$ & $6.18$ & $15.59$\\
  & $\psi_{TCVaR}$& $2.78$ & $4.07$ & $10.33$ & $26.00$\\ \hline \hline
$\nu=2.25$  & $\psi_{TVaR}$& $1.29$ & $2.00$ & $5.68$ & $15.85$\\
  & $\psi_{TCVaR}$& $2.40$ & $3.65$ & $10.25$ & $28.54$\\ \hline \hline
\end{tabular}
\end{center}
\caption{$\sigma$-multipliers for Student T VaR and CVaR loss functions}
\end{table}
But in the tails, for the values of $\nu$ suggested by both the power-law tails and MLE estimation, there is serious danger of movements of more than $16\sigma$, peaking at a $28.5\sigma$ 
average loss in events happening less than once every 40 years. 

\subsection{Excel limitations}
The generation of such quantiles in major modelling environments {\it ought} to be a matter of routine, but the Excel program still continues to exhibit some awkwardness, even within Excel 2011. I itemized the problems with the older TINV function in a supplement to \cite{shawjcf}, still available at 

\url{http://www.mth.kcl.ac.uk/~shaww/web_page/papers/Tsupp/TINV.pdf}.

In more recent versions of Excel, there is the better and new function T.INV which at least makes {\it some} direct computations of the quantiles in the left tail more clearly accessible. If you try, for example, T.INV(u,n) and normalize to unit variance you will get answers corresponding to those in table above. However, non-integer degrees of freedom are just not managed. If you set $\nu = 3.9$ you will get the results for $\nu = 3$.  To be quite clear, Microsoft's online documenation for T.INV makes it clear that only integer degrees of freedom are managed. Associated with the variable \begin{verbatim}deg_freedom\end{verbatim} there is the following remark in the online documentation for Excel 2011 Mac edition: ``If this argument contains a decimal value, this function ignores the numbers to the right side of the decimal point.'' So far as I can determine, quantiles for strictly integer degrees of freedom are computed well and deep into the tail. This is of course sufficient within the use of the T as a small sample statistic, where the degrees of freedom is an integer, but does not go far enough for our purposes, where non-integer degrees of freedom are commonplace.

\section{Risk visualization}
The functions $\psi_{DR}(u)$ for $D\in\{G,T\}$ and $R\in\{V,C\}$ characterize the loss function for  the standardized zero-mean unit-variance case of Gaussian/T and VaR/CVaR. It it is important to do the unit-variance normalization for the T as otherwise in making comparisons one is not working on a like-with-like basis. A comparison of the $\psi$-functions isolates the effect of the distributional choice on the loss. There may of course be additional effects due to a shift in the mean and an explosion in the variance, but we can now assess the effect on our perception of risk of changing the distribution alone to one that is much more realistic. At the same time we can parametrize the effect by the variable $u$, which gives us the event frequency.

In Figure 1 we show a screen shot from a dynamic visualization\footnote{These were first shown by the author at the NERC and the Knowledge Transfer Networks for Industrial Mathematics and for Financial Services workshop on Uncertainty and Risk Visualisation, Lloyd's of London, May 2010.} of the VaR and CVaR as a function of the power-law in the tails, modelled as a T distribution. The blue curves are Gaussian, green a T with $\nu=4$, and red a T with $\nu = 2.25$. There are two curves of each colour, with the upper curve being the CVaR. The loss is given in multiples of the standard deviation, so the curves are our $\psi$-functions. The horizontal axis gives the logarithmic frequency. Note that in the most extreme case we can get to $30\sigma$ events on sensible time-scales, whereas in a Gaussian model even a $5\sigma$ event needs around 100 centuries. 

\begin{figure}[hbt]
\centering
\includegraphics[scale=0.7]{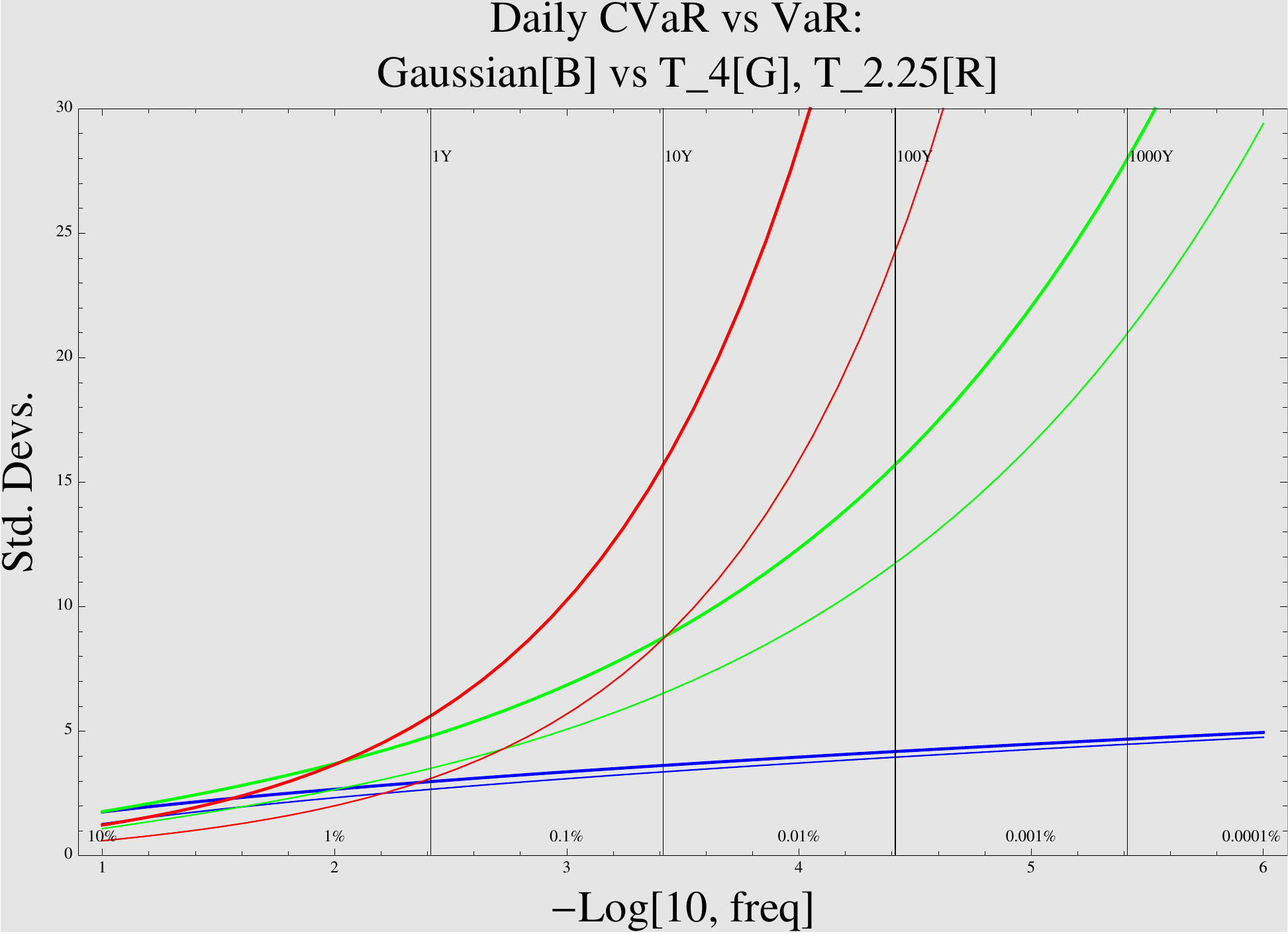}
\caption{Loss functions for Gauss (blue), $T_{2.25}$ (red), $T_4$ (green): CVaR (upper) \& VaR (lower)}
\end{figure}

\section{Portfolio Optimization}
The VaR and CVaR in any of these models is given by a risk function of the form
\begin{equation}
{\rm Risk}(u, w_i) = -\sum_{i=1}^N\mu_i w_i + \psi(u) \sqrt{\sum_{i,j}C_{ij}w_i w_j}
\end{equation}
All that has changed by the introduction of the Student T model is the setting of a new scale for the $\sigma$-weighting, which can now be dramatically higher when one worries about lower-frequency events. The values for some cases of interest can just be read off Table 2. By way of illustration, here are some sample problems:

\smallskip

\noindent
{\bf Problem 1} Optimize Gaussian VaR at at one in 40 day level: Minimize
\begin{equation}
 -\sum_{i=1}^N\mu_i w_i + 1.96 \sqrt{\sum_{i,j}C_{ij}w_i w_j}
\end{equation}

\noindent
{\bf Problem 2} Optimize Gaussian CVaR at one in 40 year level: Minimize
\begin{equation}
 -\sum_{i=1}^N\mu_i w_i + 3.96 \sqrt{\sum_{i,j}C_{ij}w_i w_j}
\end{equation}
Note how in problems 1 and 2 the simultaneous switch to CVaR and switching days to years has only doubled the $\sigma$-weighting. On the other hand...

\smallskip

\noindent
{\bf Problem 3} Optimize Student T CVaR at one in 40 year level, with a cubic tail model: Minimize
\begin{equation}
 -\sum_{i=1}^N\mu_i w_i + {\bf 19.3} \sqrt{\sum_{i,j}C_{ij}w_i w_j}
\end{equation}
So it is intuitively clear that these problems are intuitively driving us to increase the risk-aversion. Once we have the VaR and CVaR optimizations in this form, they may be fed to a wide variety of simple optimizers to find the optimal weights. It is self-evident that the impact of the remodelling of the input distribution is to drive the risk-optimal solution closer to one of pure minimum variance as $\psi$ becomes larger, but all the points are on the same efficient frontier.

\begin{figure}[hbt]
\centering
\includegraphics[scale=0.66]{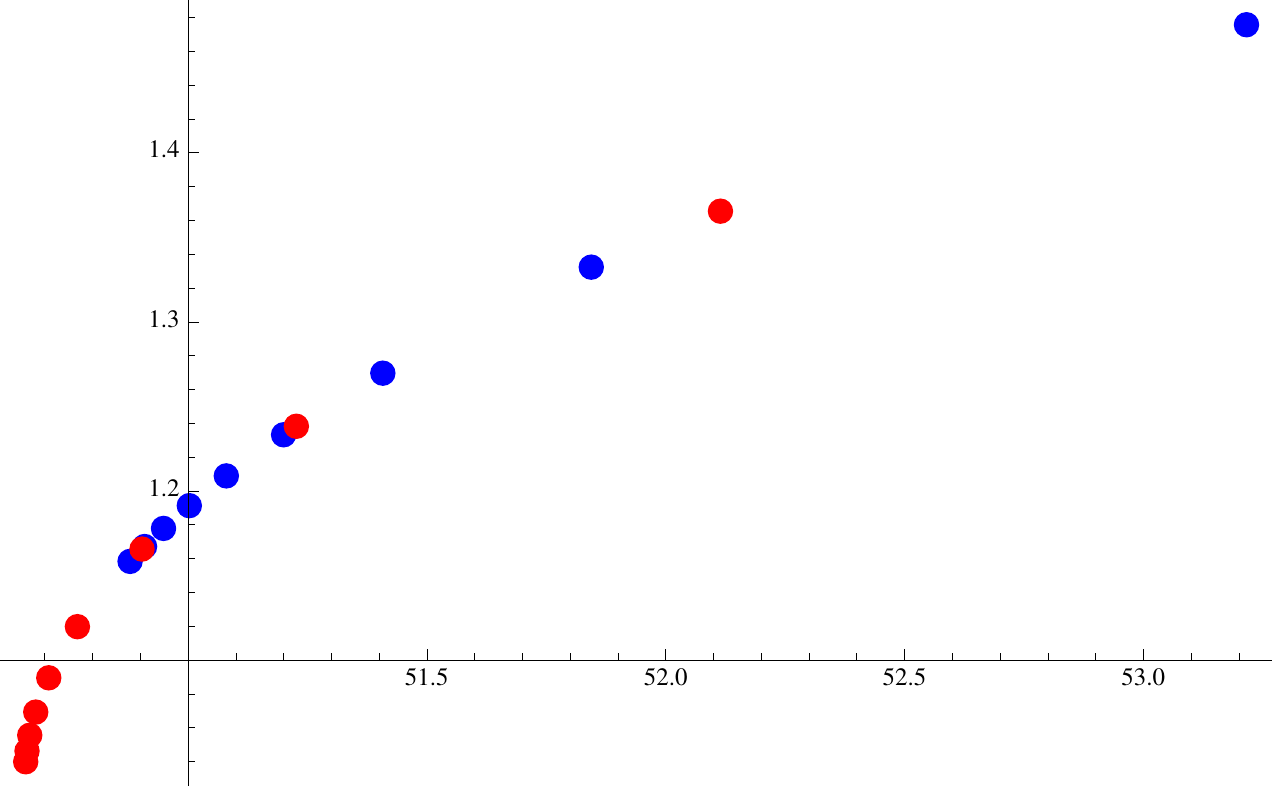}
\caption{Sampling an efficient frontier with optima for Gaussian VaR (blue) and Student T CVaR (red)}
\end{figure}

This may easily be illustrated using any simple optimizer. With values of $u$ in the form $u = 10^{-x}$, with $x$ in the range $[1, 5]$ in steps of a half, a simple 3D test problem was optimized first under Gaussian VaR and then under Student T CVaR with $\nu = 3$. The optima are shown in the figure in blue for Gaussian VaR and red for T CVaR, and an obvious pattern is obtained with return (vertical) plotted against variance (horizontal). The covariance matrix and expected returns are the same in both cases. A {\it Mathematica} notebook with the calculation details is available on request, but Figure 2 tells the obvious story. These results have also been compared with the random portfolio optimization method described in \cite{shawssrn} and good agreement has been found with a sample of portfolio returns of size $100,000$ for larger $u$ - values.

\section{Summary and Comments}
In this paper we have given analytical formulae for the VaR and CVaR of a system where the returns are characterized by a Student T distribution. The risks of a given loss measured in standard deviations are massively bigger than in the Gaussian case and may be computed simply. A switch from VaR to CVaR is in practice vacuous if the system remains Gaussian, but may make a large difference with power-law tails in the returns. The problem of optimizing VaR and CVaR is reduced to a trivial moment-based optimization and the outcome is to make a simple enhancement of the weighting of the variance component. Optima in a fat-tailed world are explicitly driven towards the pure minimum variance configuration. 

None of these results should be used to infer that better investigation of tail statistics need not  be done. Here maximum-likelihood estimates of historical distributions are being extrapolated into statements about tails, and all the usual caveats apply. 

We might ask whether this kind of analysis can be done fore more general distributions? For systems with finite variance, we can always apply a scaling and translation of the quantile function such that, for the VaR of a portfolio,
\begin{equation}
VaR(u) =\mu + \sigma \tilde{Q}_0(u,\alpha,\beta,\dots)
\end{equation}
where $\tilde{Q}_0$ is a function of other parameters $\alpha,\beta,\dots$ beyond the mean and variance, and satisfies
\begin{equation}
\int_0^1 \tilde{Q}_0(u,\alpha,\beta,\dots) = 0 \ ,\ \ \  \int_0^1 \tilde{Q}_0^2(u,\alpha,\beta,\dots) = 1\ .
\end{equation}
The problem in general is that $\alpha,\beta,\dots$ will themselves be functions of the weights. In the case of the Student T in standard multivariate form we have homogenity of the system in that the same value of the single extra parameter $\nu$ applies to all the assets and to any portfolio built from them by linear combination. This is what allows us to work with such a system using simple analytics. Similar constructions could be done with any conditionally Gaussian form with a homogeneous randomization of the variance.

The determination of $\nu$ is something that is probably best done, at least initially, by maximum likelihood estimation. The temptation to use a simple kurtosis model should be resisted. This is for the simple reason that the total kurtosis is given by
\begin{equation}
\kappa_{tot} = 3 + \frac{6}{\nu-4}
\end{equation}
and is undefined for $\nu < 4$, which is a domain of considerable interest given the data studies. 

Finally we note that is likely that other risk functions will admit a similar decomposition to simple mean-s.d. form for the case of the Student T.

\section*{Appendix: Tail series coefficients for the T-quantile}
\begin{equation}
\begin{split}
d_1 &= 1\\
d_2& = -\frac{1}{\nu +2}\\
d_3 &=-\frac{(\nu -2) (\nu +3)}{2 (\nu +2)^2 (\nu +4)}\\
d_4 &= -\frac{(\nu -2) \left(\nu ^3+6 \nu ^2+2 \nu -18\right)}{3 (\nu +2)^3 (\nu +4) (\nu +6)}\\
d_5 &= -\frac{(\nu -2) (\nu +5) \left(6 \nu ^5+59 \nu ^4+95 \nu ^3-284 \nu ^2-380 \nu +576\right)}{24 (\nu +2)^4 (\nu
   +4)^2 (\nu +6) (\nu +8)}\\
   d_6 &= -\frac{(\nu -2) (\nu +3) \left(2 \nu ^7+37 \nu ^6+192 \nu ^5+26 \nu ^4-1430 \nu ^3-48 \nu ^2+3576 \nu
   -2400\right)}{10 (\nu +2)^5 (\nu +4)^2 (\nu +6) (\nu +8) (\nu +10)}
\end{split}
\end{equation}
The associated recurrence for making more terms is given in Section 3.4 of \cite{ejam}. Precision of better than $10^{-9}$ on the region $0<u\leq0.025$ is obtained for $2\leq \nu \leq 18$ with e.g. 40 terms and the answers are then double precision quality for $\nu < 6$.


\begin{thebibliography}{10}

\bibitem{ackweb}
{\sc P. J. Acklam}  An algorithm for computing the inverse normal cumulative distribution function,
\url{http://home.online.no/~pjacklam/notes/invnorm}

\bibitem{epjb}
{\sc W. Breymann, D.R. L\"{u}thi, E. Platen}, 2009,  Empirical behavior of a world stock index from intra-day to monthly time scales, {\em European Physics Journal} B, Vol. 71 (4), 511-522.

\bibitem{fama}
{\sc E.F. Fama}, 1965. The behaviour of stock prices, {\em Journal of Business}, 37, 34-105. 


\bibitem{platent}
{\sc K. Fergusson and E. Platen}, On the Distributional Characterization of daily Log-returns of a World Stock Index, Applied Mathematical Finance, 13 (1), 19-38, 2006

\bibitem{hill}
{\sc G.W. Hill}, 1970, Algorithm 396, Student's t-quantiles, {\em Communications of the ACM}, 13  (10), 619-620. 

\bibitem{stanley}
{\sc X. Gabaix, P. Gopikrishnan, V. Plerou, and H. E. Stanley}, 2003, A Theory of Power-Law Distributions in Financial Market Fluctuations, {\em Nature} 423, 267-270. 

\bibitem{mandel}
{\sc B. Mandelbrot}, 1963. The variation of certain speculative prices, {\em Journal of Business} 36, 394-419. 


\bibitem{markow1} 
{\sc H.~M. Markowitz}, 
{\em Portfolio Selection},  Journal of Finance, 7, pp.~77--91 (1952). 

\bibitem{markow2} 
{\sc H.~M. Markowitz}, 
{\em The Optimization of a Quadratic Function Subject to Linear Constraints},  Naval Research Logistics Quarterly 3, pp.~111-33, 1956. 

\bibitem{markow3}
{\sc H.~M. Markowitz}, {\em Portfolio Selection: Efficient Diversification of Investments}. Cowles Foundation Monograph No. 16. New York: Wiley, 1959.

\bibitem{cvaropt}
{\sc R.T. Rockafellar, S. Uryasev},  {\em Optimization of Conditional Value-At-Risk}. The Journal of Risk, Vol. 2, No. 3, 2000, 21-41


\bibitem{shawjcf}
{\sc W.~T. Shaw}, 2006, {\em Journal of Computational Finance}

\bibitem{shawssrn}
{\sc W.~T. Shaw}, 2010, {\em Monte Carlo Portfolio Optimization for General Investor Risk-Return Objectives and Arbitrary Return Distributions: A Solution for Long-Only Portfolios}. \url{http://ssrn.com/abstract=1680224}

\bibitem{jmva}
{\sc W.~T. Shaw, K.~T.~A. Lee}, 2008, Bivariate Student t distributions with variable marginal degrees of freedom and independence, {\em Journal of Multivariate Analysis}, 99, 1276-1287.

\bibitem{ejam}
{\sc G. Steinbrecher, W.~T.Shaw}, 2008, Quantile Mechanics, {\em European Journal of Applied Mathematics}, 19(2), 87-112.

\bibitem{student}
{\sc Student}, a.k.a. {\sc W.S. Gosset}, 1908, The probable error of a mean. {\em Biometrika} 6, 1-25.

\bibitem{taleb}
{\sc N.N. Taleb}, 2008, Finiteness of Variance is Irrelevant in the Practice of Quantitative Finance, \url{http://ssrn.com/abstract=1142785}.

\bibitem{wichura}
{\sc Wichura, M.J.}, Algorithm AS 241: The Percentage Points of the Normal Distribution. {\em Applied Statistics}, 37, 477-484, 1988.


\end{thebibliography}
\end{document}